\documentclass[aps,prd,floatfix,superscriptaddress,preprintnumbers]{revtex4}
\usepackage{amssymb}
\usepackage{amsmath}
\usepackage{amsfonts}
\usepackage{epsfig}             
\usepackage{graphicx}
\usepackage{tabularx}
\usepackage{color}
\newcommand{\seq}{\begin{subequations}}
\newcommand{\sen}{\end{subequations}}
\newcommand{\eq}{\begin{eqnarray}}
\newcommand{\en}{\end{eqnarray}}

\def\shiftdown#1{#1\llap{\lower.04ex\hbox{#1}}}

\begin{document}

\title{$\gamma N \to N^*(1535)$ transition in soft-wall AdS/QCD} 

\author{Thomas Gutsche}
\affiliation{Institut f\"ur Theoretische Physik,
Universit\"at T\"ubingen, 
Kepler Center for Astro and Particle Physics,  
Auf der Morgenstelle 14, D-72076 T\"ubingen, Germany}
\author{Valery E. Lyubovitskij} 
\affiliation{Institut f\"ur Theoretische Physik,
Universit\"at T\"ubingen, 
Kepler Center for Astro and Particle Physics,  
Auf der Morgenstelle 14, D-72076 T\"ubingen, Germany}
\affiliation{Departamento de F\'\i sica y Centro Cient\'\i fico
Tecnol\'ogico de Valpara\'\i so-CCTVal, Universidad T\'ecnica
Federico Santa Mar\'\i a, Casilla 110-V, Valpara\'\i so, Chile}
\affiliation{Department of Physics, Tomsk State University,
634050 Tomsk, Russia}
\author{Ivan Schmidt}
\affiliation{Departamento de F\'\i sica y Centro Cient\'\i fico
Tecnol\'ogico de Valpara\'\i so-CCTVal, Universidad T\'ecnica
Federico Santa Mar\'\i a, Casilla 110-V, Valpara\'\i so, Chile}

\date{\today}

\begin{abstract}

We present a study of the $N^*(1535)$ 
resonance electroexcitation in a soft-wall AdS/QCD model. 
Both the transverse $A_{1/2}^p$ and longitudinal 
$S_{1/2}^p$ helicity amplitudes are calculated resulting in good agreement 
with data and with the MAID parametrization. 

\end{abstract}

\maketitle

\section{Introduction}

The study of electromagnetic transitions between the nucleon and 
its resonances is an important area of research in hadronic physics,
because it can provide essential information about the structure and
basic properties of the involved hadrons~\cite{Devenish:1975jd,%
Aznauryan:2011qj,Aznauryan:2009mx,Aznauryan:2012ec,Mokeev:2015lda,%
Tiator:2011pw,Stajner:2017fmh}. 
In fact, right now there are recent experiments
at JLab~\cite{Aznauryan:2009mx,Aznauryan:2012ec,Mokeev:2015lda} and
at MAMI~\cite{Tiator:2011pw,Stajner:2017fmh} aiming for a precise
determination of the electrocouplings of nucleon resonances and nucleons,
which should also be complemented by theoretical studies of these physical
properties (for a review see, e.g.,
Refs.~\cite{Aznauryan:2011qj,Tiator:2011pw}). 

In the present paper we continue our study of nucleon resonances in soft-wall 
AdS/QCD. An advantage of this approach is that it contains the correct
power scaling description of form factors and helicity amplitudes 
at large~$Q^2$~\cite{Brodsky:1973kr}, but also provides good agreement with data
at low and intermediate $Q^2$. 
The preceding study of nucleon resonances in AdS soft-wall approaches focused
on the nucleon-Roper transition,
see Refs.~\cite{deTeramond:2011qp}-\cite{Gutsche:2019jzh}.      
First, in Ref.~\cite{deTeramond:2011qp} the Dirac form factor, which 
defines the electromagnetic nucleon-Roper transition, was predicted 
in holographic light-front QCD.
In Ref.~\cite{Gutsche:2012wb} we proposed and extended the formalism
to the description of all nucleon resonances with adjustable quantum
numbers in soft-wall AdS/QCD. In a first application we looked at
a comprehensive description of Roper-nucleon transition properties,
including form factors, helicity amplitudes and charge radii. 
In Ref.~\cite{Gutsche:2017lyu} we showed that the description of 
the electromagnetic form factors of the nucleon and of the Roper resonance
can be sufficiently improved using an extended version of the effective
action of soft-wall AdS/QCD. This was achieved by including additional
nonminimal terms into the action consistent with gauge invariance,
finally resulting in important 
contributions to the momentum dependence of the form factors and helicity 
amplitudes. Moreover, in Ref.~\cite{Gutsche:2019jzh} we presented a 
description of electromagnetic properties of the nucleon and the Roper
at small finite temperatures using the formalism developed
in Ref.~\cite{Gutsche:2019blp}.  
In the present manuscript we extend our formalism, previously developed
for the study of the Roper resonance, to the negative-parity state $N^*(1535)$. 

The electromagnetic nucleon to $N^*(1535)$ transition  has been studied 
in several theoretical approaches. 
In Ref.~\cite{Konen:1989jp} the electromagnetic form factors and the
helicity amplitudes of the $N^*(1535)$ excitation
were calculated in a constituent quark model and formulated in
light front dynamics. The electroexcitation of the $N^*(1535)$ resonance 
has also been studied in Ref.~\cite{Aznauryan:2004jd}, using two methods: 
dispersion relations and the isobar model. 
The unitary isobar model MAID has been developed to analyze the world data 
of pion photoproduction and electroproduction, including the $N^*(1535)$ 
resonance~\cite{Drechsel:2007if}. 
The evaluation of the electromagnetic helicity form factors 
for the electroproduction of the $N^*(1535)$ resonance, considered as 
a dynamically generated resonance, has been addressed 
in Ref.~\cite{Jido:2007sm}. 
In Ref.~\cite{Braun:2009jy} the $N^*(1535)$ resonance electroproduction
has been studied in the framework of light-cone sum rules, which combine 
perturbative relations and duality. In Ref.~\cite{Aliev:2013dxa}
the $\gamma^\ast N \to N(1535)$ transition form factors were analyzed in 
the context of QCD sum rules. 
In Ref.~\cite{Ramalho:2011ae} this resonance was studied using a relativistic 
constituent quark model. In Refs.~\cite{Ramalho:2011fa,An:2008xk}   
several methods (semirelativistic approximation, empirical parametrization, 
a new mechanism for $\gamma \to q\bar q$ coupling, inclusion 
of the lowest lying pentaquark components $qqqq\bar q$), have been proposed 
for a better understanding of the $\gamma N \to N^*(1535)$ transition. 
Most of the approaches (except in the MAID~\cite{Drechsel:2007if} analysis)
fail in the description of the longitudinal $S_{1/2}$ amplitude 
in the low $Q^2$ domain. In Refs.~\cite{Obukhovsky:2013fpa,Obukhovsky:2019aa} 
a unified description of the electroexcitation of the Roper and $N(1535)$
resonances has been done in the light-front quark model. A reasonable
description of the helicity amplitudes at intermediate and high $Q^2$ has 
been obtained.

The main aspect of the present paper is that we can indicate
a mechanism which leads to reasonable results for the helicity amplitudes
also in the low-$Q^2$ regime: inclusion of the minimal coupling of the
nucleon and the $N^*(1535)$ Fock components with the same twist dimension. 
The nucleon has orbital momentum $L=0$, while the negative-parity resonance
$N^*(1535)$ has $L=1$. Therefore, 
the leading twists for the nucleon and the $N^*(1535)$ are $\tau = 3$ and 
$\tau = 4$, respectively. Then the only possibility is that the leading twists 
of the nucleon and $N^*(1535)$ couple with the photon via a nonminimal
coupling, since the minimal one is forbidden by current conservation.  
But we find that the minimal coupling between the nucleon and the $N^*(1535)$ 
is possible for equal twists, which means that the 
leading minimal electromagnetic coupling between nucleon and $N^*(1535)$ 
occurs for $\tau_N = \tau_{N^*(1535)} = 4$. Inclusion of this particular coupling 
helps to improve the description of both helicity amplitudes $A_{1/2}$  
and $S_{1/2}$ at low $Q^2$. 

The paper is organized as follows. 
In Sec.~II we briefly discuss our formalism. 
In Sec.~III we present the analytical calculation and the numerical analysis 
of electromagnetic form factors and helicity amplitudes of the 
nucleon-$N(1535)$ transition.  
Finally, Sec.~IV contains our summary and conclusions.

\section{Formalism}

In this section we briefly review our approach~\cite{Gutsche:2011vb,%
Gutsche:2012bp,Gutsche:2012wb,Gutsche:2017lyu}. 
We start with the definition of the conformal Poincar\'e metric 
\eq
g_{MN} \, x^M x^N = \epsilon^a_M \, \epsilon^b_N \, 
\eta_{ab} \, x^M x^N = \frac{1}{z^2} \, (dx_\mu dx^\mu - dz^2),
\en 
where $\epsilon^a_M = \delta^a_M/z$ is the              
vielbein, and we define $g = |{\det}(g_{MN})| = 1/z^{10}$ as the magnitude 
of the determinant of $g_{MN}$. 

The soft-wall AdS/QCD action $S$ for the nucleon 
$N = (p,n)$ and the $N^*(1535)$ resonance 
[in the following we use the notation $N^* = (N^*_p,N^*_n)$] 
including photons is constructed in terms of the dual spin-$1/2$ fermion and 
vector fields. These fields have constrained (confined) dynamics in AdS space 
due to the presence of a background field --- dilaton field 
$\varphi(z) = \kappa^2 z^2$, where $\kappa$ is its scale parameter. 

The action $S$ contains a free part $S_0$, describing the 
confined dynamics of AdS fields, and an interaction part $S_{\rm int}$, 
describing the interactions of fermions with the vector field 
[below, for simplicity, we only display the coupling of $N$ and $N^*(1535)$ 
to the vector field]    
\eq\label{actionS}
S   &=& S_0 + S_{\rm int}\,, \nonumber\\[3mm]
S_0 &=& \int d^4x dz \, \sqrt{g} \, e^{-\varphi(z)} \,
\biggl\{ {\cal L}_N(x,z) + {\cal L}_{N^*}(x,z)
+ {\cal L}_V(x,z)
\biggr\} \,, \nonumber\\[3mm]
S_{\rm int} &=& \int d^4x dz \, \sqrt{g} \, e^{-\varphi(z)} \,
{\cal L}_{VNN^*}(x,z) \,. 
\en
${\cal L}_N$, ${\cal L}_{N^*}$, ${\cal L}_V$ and
${\cal L}_{VNN^*}$ are the free and
interaction Lagrangians,
respectively, and are written as 
\eq\label{actionS2}
{\cal L}_N(x,z) &=&  \sum\limits_{i=+,-; \,\tau} \, c_\tau^N \,
\bar\psi^N_{i,\tau}(x,z) \, \hat{\cal D}_i(z) \, \psi^N_{i,\tau}(x,z)
\,, \nonumber\\[3mm]
{\cal L}_{N^*}(x,z) &=&  \sum\limits_{i=+,-; \,\tau} \, c_{\tau+1}^{N^*} \,
\bar\psi^N_{i,\tau+1}(x,z) \, \hat{\cal D}_i(z) \, \psi^{N^*}_{i,\tau+1}(x,z)
\,, \nonumber\\[3mm]
{\cal L}_V(x,z) &=& - \frac{1}{4} V_{MN}(x,z)V^{MN}(x,z)\,, 
\nonumber\\[3mm]
{\cal L}_{VNN^*}(x,z) &=&
\sum\limits_{i=+,-; \,\tau} \, \biggl[ c_{\tau+1}^{N^*N} \,
\bar\psi_{i,\tau+1}^{N^*}(x,z) \, \hat{\cal V}^{N^*N}_{i,m}(x,z) \,
\psi_{i,\tau+1}^N(x,z) \nonumber\\
&+&  d_\tau^{N^*N} \,
\bar\psi_{i,\tau+1}^{N^*}(x,z) \, \hat{\cal V}^{N^*N}_{i,nm}(x,z) \,
\psi_{i,\tau}^N(x,z) \biggr]
+ {\rm H.c.}\,.
\en
Here $\tau$ runs from 3. 
We thereby introduce the shortened notation
\eq
\hat{\cal D}_\pm(z) &=&  \frac{i}{2} \Gamma^M
\! \stackrel{\leftrightarrow}{\partial}_{_M} - \frac{i}{8}
\Gamma^M \omega_M^{ab} [\Gamma_a, \Gamma_b]
\, \mp \,  (\mu + U_F(z))\,, \nonumber\\[3mm]
\hat{\cal V}^{N^*N}_{\pm,m}(x,z)  &=&  Q \, \Gamma^M  V_M(x,z)\,, 
\nonumber\\[3mm] 
\hat{\cal V}^{N^*N}_{\pm,nm}(x,z) &=& \pm \,
\frac{i}{4} \, \eta_V^H \,  [\Gamma^M, \Gamma^N] \, V_{MN}(x,z) 
+ \zeta_V^{N^*N} \, z \, \Gamma^M  \, \partial^N V_{MN}(x,z)
\,,
\en
where 
$\mu$ is the five-dimensional mass of the spin-$\frac{1}{2}$ AdS      
fermion with $\mu = 3/2 + L$ ($L$ is the orbital angular momentum);      
$U_F(z) = \varphi(z)$ is the dilaton potential; 
$Q = {\rm diag}(1,0)$ is the nucleon ($N^*$) charge matrix;  
$V_{MN} = \partial_M V_N - \partial_N V_M$ is                            
the stress tensor for the vector field; 
$\omega_M^{ab} = (\delta^a_M \delta^b_z - \delta^b_M \delta^a_z)/z$ is 
the spin connection term; $\sigma^{MN} = [\Gamma^M, \Gamma^N]$ 
is the commutator of the Dirac matrices in AdS space, which are defined as  
$\Gamma^M = \epsilon^M_a \Gamma^a$ and 
$\Gamma^a = (\gamma^\mu, -i \gamma^5)$. The subscripts $m$ and $nm$ 
in the vector matrix $\hat{\cal V}^{N^*N}_{\pm,m/nm}(x,z)$ refer to 
minimal $(m)$ or nonminimal $(nm)$ couplings, respectively.   
As was pointed out in the Introduction, due to gauge invariance 
the minimal coupling between the nucleon and 
the $N^*$ is only possible for the same twist, while for the nonminimal 
coupling it is not constrained and we include the coupling between 
nucleon with twist $\tau_N = 3,4,5$ and the $N^*(1535)$ resonance with the 
twist $\tau_{N^*} = \tau_N+1 = 4,5,6$. 

The action~(\ref{actionS}) is constructed in terms of the 
5D AdS fermion fields $\psi^N_{\pm,\tau}(x,z)$, 
$\psi^{N^*}_{\pm,\tau}(x,z)$ and the vector field $V_M(x,z)$. 
Fermion fields are duals to the left- and 
right-handed chiral doublets of the nucleon and the $N^*(1535)$ resonance with
${\cal O}^L = (B_1^L, B_2^L)^T$ and ${\cal O}^R = (B_1^R, B_2^R)^T$ 
where $B_1 = p, N^*_p$ and $B_2 = n, N^*_n$. These fields  
are in the fundamental representations of the chiral $SU_L(2)$ 
and $SU_R(2)$ subgroups and are holographic analogs of the nucleon $N$ 
and $N^*(1535)$ resonance, respectively. 

The 5D AdS fields $\psi^B_{\pm,\tau}(x,z)$ are products of the left/right 
4D spinor fields 
\eq\label{psi_expansion_Nucleon}
\psi^{L/R}_N(x) = \frac{1 \mp \gamma^5}{2} \, \psi(x)
\en
for the nucleon, and 
\eq\label{psi_expansion_Nstar}
\psi^{L/R}_{N^*}(x) = \gamma^5 \, \frac{1 \mp \gamma^5}{2} \, \psi(x) 
= \mp \frac{1 \mp \gamma^5}{2} \, \psi(x) 
\en 
for the $N^*(1535)$ resonance 
with spin $1/2$, and the bulk profiles 
\eq\label{psi_expansion1}
F^{L/R}_{\tau}(z) = z^2 \, f^{L/R}_{\tau}(z),
\en 
with twist $\tau$, which depend on the holographic (scale) variable $z$: 
\eq\label{psi_expansion2}
\psi^N_{\pm,\tau}(x,z) &=& \frac{1}{\sqrt{2}} \,
\left[
\pm \psi^{L}_N(x) \ F^{L/R}_{\tau}(z)
+   \psi^{R}_N(x) \ F^{R/L}_{\tau}(z)\right]\,, \nonumber\\
\psi^{N^*}_{\pm,\tau}(x,z) &=& \frac{1}{\sqrt{2}} \,
\left[
\mp \psi^{L}_{N^*}(x) \ F^{L/R}_{\tau}(z)
+   \psi^{R}_{N^*}(x) \ F^{R/L}_{\tau}(z)\right]\,,
\en
where
\eq\label{fL_fR} 
f^L_{\tau}(z) &=& \sqrt{\frac{2}{\Gamma(\tau)}} \, \kappa^{\tau} \,
z^{\tau - 1/2} \, e^{-\kappa^2 z^2/2}\,, \nonumber\\
f^R_{\tau}(z)  &=& \sqrt{\frac{2}{\Gamma(\tau-1)}} \, \kappa^{\tau-1} \,
z^{\tau - 3/2} \, e^{-\kappa^2 z^2/2}\,. 
\en
The nucleon is identified as the ground state with $n=L=0$ while the $N^*(1535)$ resonance as the
first orbitally excited state with $n=0$ and $L=1$.  
In the case of the vector field we work in the axial gauge
$V_z = 0$ and perform a Fourier transformation of 
the vector field $V_\mu(x,z)$ with respect to the Minkowski coordinate
\eq\label{V_Fourier}
V_\mu(x,z) = \int \frac{d^4q}{(2\pi)^4} e^{iqx} V_\mu(q) V(q,z)\,. 
\en

We can then derive an equation of motion for the vector bulk-to-boundary
propagator $V(q,z)$ dual to the $q^2$-dependent electromagnetic current
\eq
\partial_z \biggl( \frac{e^{-\varphi(z)}}{z} \,
\partial_z V(q,z)\biggr) + q^2 \frac{e^{-\varphi(z)}}{z} \, 
V(q,z) = 0 \,. 
\en 
The solution of this equation in terms of the
gamma $\Gamma(n)$ and Tricomi $U(a,b,z)$ functions reads
\eq
\label{VInt_q}
V(q,z) = \Gamma\Big(1 - \frac{q^2}{4\kappa^2}\Big)
\, U\Big(-\frac{q^2}{4\kappa^2},0,\kappa^2 z^2\Big) \,.
\en
In the Euclidean region ($Q^2 = - q^2 > 0$)
it is convenient to use the integral
representation for $V(Q,z)$~\cite{Grigoryan:2007my}
\eq
\label{VInt}
V(Q,z) = \kappa^2 z^2 \int_0^1 \frac{dx}{(1-x)^2}
\, x^{a} \,
e^{- \kappa^2 z^2 \frac{x}{1-x} }\,,
\en
where $x$ is the light-cone momentum fraction and
$a = Q^2/(4 \kappa^2)$.

The set of parameters 
$c_{\tau}^{N}$, $c_{\tau+1}^{N^*}$, 
$c_{\tau+1}^{N^*N}$, and $d_\tau^{N^*N}$ 
induces mixing of the contributions of AdS 
fields with different twist dimensions. 
In Refs.~\cite{Gutsche:2012bp,Gutsche:2012wb} we
showed that the parameters $c_\tau^N$ and $c_{\tau+1}^{N^*}$ 
are constrained by the conditions $\sum_\tau \, c_\tau^N = 1$ 
and $\sum_\tau \, c_{\tau+1}^{N^*} = 1$ 
in order to get the correct normalization of the kinetic term
$\bar\psi(x)i\!\!\not\!\!\partial\psi(x)$
of the four-dimensional spinor field. This condition is also
consistent with electromagnetic gauge invariance. 
Therefore, the nucleon and $N^*(1535)$ masses can be identified with 
following the expressions~\cite{Gutsche:2012bp,Gutsche:2012wb}
\eq\label{Matching1}
M_N = 2 \kappa \sum\limits_\tau\, c_\tau^N\, \sqrt{\tau - 1}
\,, \quad\quad  
M_{N^*} &=& 2 \kappa 
\sum\limits_\tau \, c_{\tau+1}^{N^*}\, \sqrt{\tau}\,. 
\en
As in the previous case of the
nucleon and the Roper resonance we restrict our calculation to the three 
leading twist contributions to the $N^*(1535)$ mass 
$(\tau_{N^*} = 4,5,6)$. 
With the condition $\sum_\tau \, c_{\tau+1}^{N^*} = 1$ 
only two parameters are linearly independent. One of the possible solutions  
fixing the central value of the $N^*(1535)$ mass of 1510 MeV~\cite{PDG18} 
reads: $c_4^{N^*} = 0.82$, $c_5^{N^*} = -0.63$, 
and $c_6^{N^*} = 1 - c_4^{N^*} - c_5^{N^*} = 0.81$.   

The baryon form factors are determined analytically
using the bulk profiles of fermion fields
and the bulk-to-boundary propagator $V(Q,z)$ of 
the vector field (for exact expressions see the next section). 
The calculational technique was already described in detail
in Refs.~\cite{Gutsche:2012bp,Gutsche:2012wb,Gutsche:2017lyu}.
The parameter $\kappa = 383$ MeV is universal and was fixed 
in previous studies (see, e.g., Refs.~\cite{Gutsche:2012bp,Gutsche:2012wb}). 
The other parameters are fixed from a fit to the helicity amplitudes of 
the $\gamma N \to N^*(1535)$ transition: 
\eq 
& &c_4^{N^*N} =  25.52\,, \quad 
   c_5^{N^*N} = -26.90\,, \nonumber\\
& &d_3^{N^*N} = -1.89\,, \quad 
   d_4^{N^*N} =  5.64\,, \quad 
   d_5^{N^*N} = -3.58\,, \nonumber\\
& &\eta_V^{N^*N}  =  4.28\,, \quad  
   \zeta_V^{N^*N} = -0.47\,. 
   \en
   
\section{Electromagnetic form factors and helicity 
amplitudes of the $\gamma N \to N^*(1535)$ transition}

The electromagnetic form factors of the $\gamma N \to N^*(1535)$ transition 
are defined, due to Lorentz and gauge invariance,
by the following matrix element 
\eq\label{matrix_elements}
M^\mu(p_1\lambda_1,p_2\lambda_2) &=& \bar u_{N^*}(p_1\lambda_1)
\biggl[ \gamma^\mu_\perp \, F_1^{N^*N}(-q^2) 
+ i \sigma^{\mu\nu} \frac{q_\nu}{M_+} \, F_2^{N^*N}(-q^2) 
\, \biggr] \, \gamma^5 \, u_{N}(p_2\lambda_2)\, .
\en
We have $u_{N^*}(p_1\lambda_1)$ and $u_{N}(p_2\lambda_2)$ which are the 
usual spin-$\frac{1}{2}$ Dirac spinors describing the 
$N^*(1535)$ resonance and nucleon,  
$M_\pm = M_{N^*} \pm M_N$, $\gamma^\mu_\perp = \gamma^\mu 
- q^\mu \not\! q/q^2$\,, 
$q = p_1 - p_2$, and $\lambda_1$, $\lambda_2$, and $\lambda$ are 
the helicities of the final, initial baryon and photon, respectively, 
with the relation $\lambda_2 = \lambda_1 - \lambda$. 
In the rest frame of the $N^*(1535)$ the four momenta of $N^*$, $N$,
photon and the polarization vector of photon are specified as:
\eq
& &p_1 = (M_1, \vec{0\,})\,, \quad 
   p_2 = (E_2, 0, 0, -|{\bf p}|)\,, \quad  
     q = (q^0, 0, 0,  |{\bf p}|)\,, \nonumber\\
& &\epsilon^\mu(\pm) = (0, - \vec{\epsilon\,}^{\pm})\,, \quad 
\vec{\epsilon\,}(\pm) = \frac{1}{\sqrt{2}} (\pm 1, i, 0)\,, \quad 
\epsilon^\mu(0) = \frac{1}{\sqrt{Q^2}}(|{\bf p}|, 0, 0, q^0)\,,  
\en
where
$ |{\bf p}| = \frac{\sqrt{Q_+ Q_-}}{2M_{N^*}}$ 
is the absolute value of the three-momentum of the nucleon or the photon. 

It is important to point out that the matrix element~(\ref{matrix_elements}) 
is manifestly gauge invariant. The form factor $F_1^{N^*N}(Q^2)$ 
vanishes at $Q^2=0$. In more detail the contribution to 
$F_1^{N^*N}(Q^2)$ from nonminimal terms of the action~(\ref{actionS})   
includes the $z$-derivative acting on the vector bulk-to-boundary 
propagator $\partial_z V(Q,z)$, which is zero at $Q^2 = 0$ 
because of $V(0,z) \equiv 1$. In the case of the minimal term its contribution 
to the $F_1^{N^*N}(Q^2)$ reads: 
\eq 
F_{1,m}^{N^*N}(Q^2) = \frac{a}{2} \, 
\sum\limits_{\tau} \, c_{\tau+1}^{N^*N} \, B(a+1,\tau+1)\,,
\en  
where $a = Q^2/(4\kappa^2)$, $B(x,y) = \Gamma(x) \Gamma(y)/\Gamma(x+y)$ 
and $\Gamma(x)$ are the beta and gamma functions. It should be clear that 
$F_{1,m}^{N^*N}(0) = 0$ at $Q^2=0$.  

Next we introduce the helicity amplitudes $H_{\lambda_1\lambda}$ which 
are related to the invariant form factors $F_i^{{\cal R}N}$ 
as (see details in
Refs.~\cite{Kadeer:2005aq,Faessler:2009xn}) 
\eq
H_{\lambda_1\lambda} = M_\mu(p_1\lambda_1,p_2\lambda_2)
\, \epsilon^{\mu}(\lambda) \,.
\en
A straightforward evaluation
gives~\cite{Devenish:1975jd,Aznauryan:2011qj,Aznauryan:2012ec,%
Kadeer:2005aq,Faessler:2009xn} 
\eq
H_{\pm\frac{1}{2}0} = \mp \sqrt{\frac{Q_+}{Q^2}} \,
\left(
F_1^{N^*N} M_-  - F_2^{N^*N} \frac{Q^2}{M_+} \right) \,, 
\quad \
H_{\pm\frac{1}{2}\pm 1} = \pm \sqrt{2 Q_+} \,
\left( F_1^{N^*N} + F_2^{N^*N}  \frac{M_-}{M_+}\right) \,, 
\en
where $Q_\pm = M_\pm^2 + Q^2$. 
In the case of the Roper-nucleon transition we also have an
additional set of helicity amplitudes $(A_{1/2}, S_{1/2})$
related to the $(H_{\frac{1}{2}0},H_{\frac{1}{2}1})$
by~
\eq
A_{1/2} = b \, H_{\frac{1}{2}1}\,, \quad
S_{1/2} = b \, \frac{|{\bf p}|}{\sqrt{Q^2}} \, H_{\frac{1}{2}0}\,,
\en
where
\eq
b = \sqrt{\frac{\pi\alpha}{M_+ M_- M_N}}
\en
and $\alpha = 1/137.036$ is the fine-structure constant. 

At $Q^2 = 0$ our predictions for the last set of helicity amplitudes
(proton channel) in the $N-N^*(1535)$ transition are 
\eq 
A_{1/2}^p(0) =  0.09 \ {\rm GeV}^{-1/2}\,, \quad\quad 
S_{1/2}^p(0) = -0.002 \ {\rm GeV}^{-1/2}\,. 
\en 

\begin{figure}[htb]
\begin{center}
\epsfig{figure=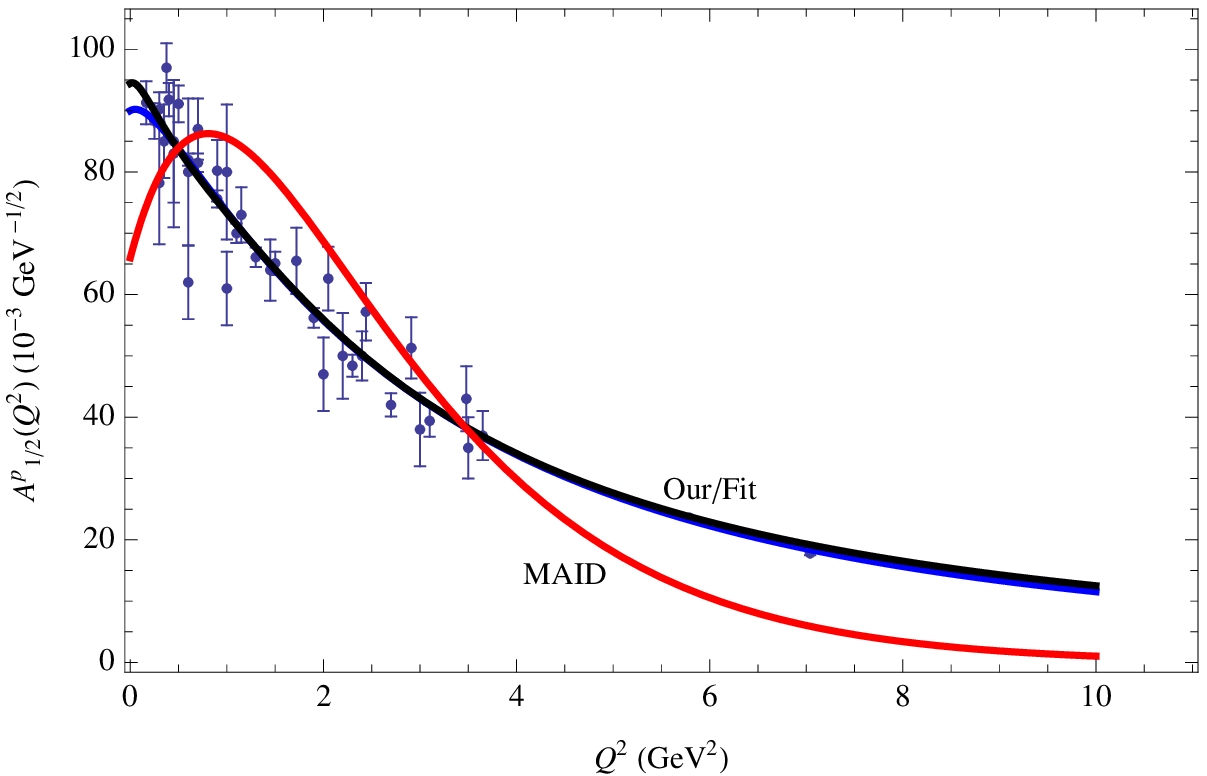,scale=.7}
\epsfig{figure=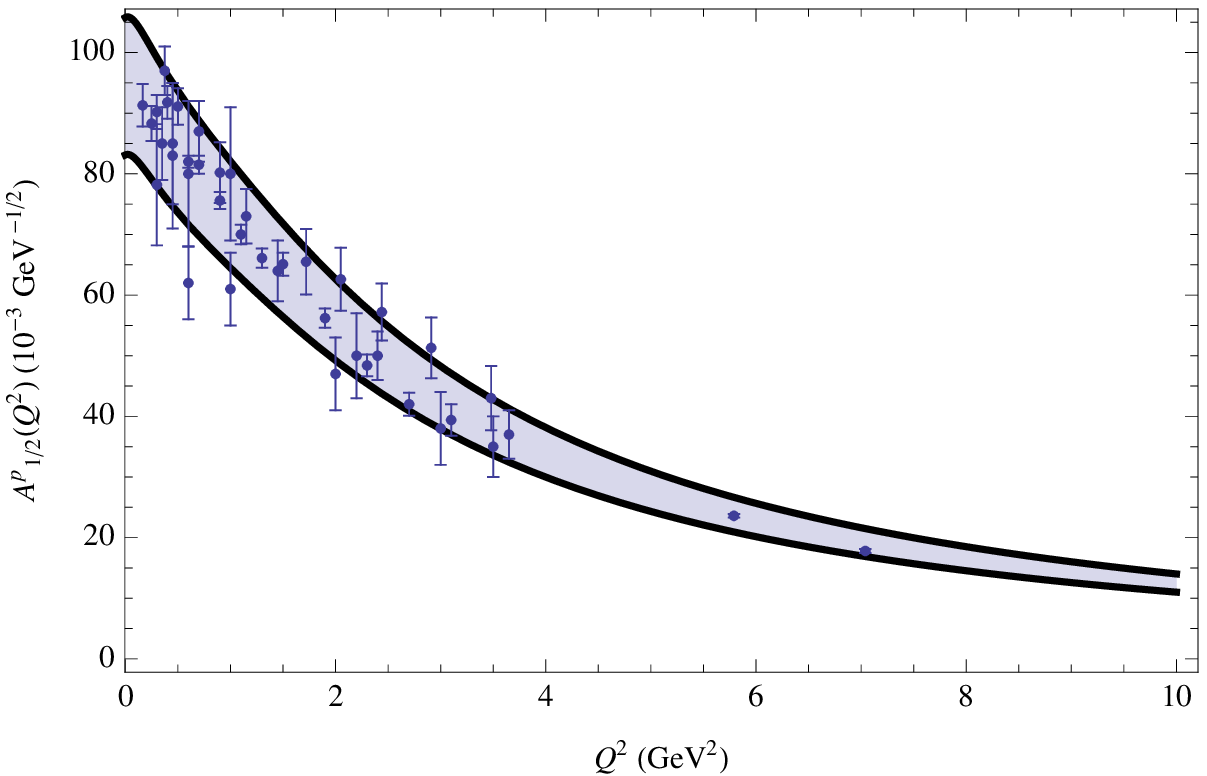,scale=.7}
\caption{Helicity amplitude $A_{1/2}^p(Q^2)$
up to $Q^2=$ 10 GeV$^2$ with data taken from CLAS. 
Left panel: Displayed are our results (Our), 
the MAID parametrization (MAID), 
and (Fit) - the parametrization of Eq.~(25). 
Right panel: Our results are shown for 
the case of variation of parameters of our approach (shaded band) 
in comparison with data. 
\label{fig1}}

\epsfig{figure=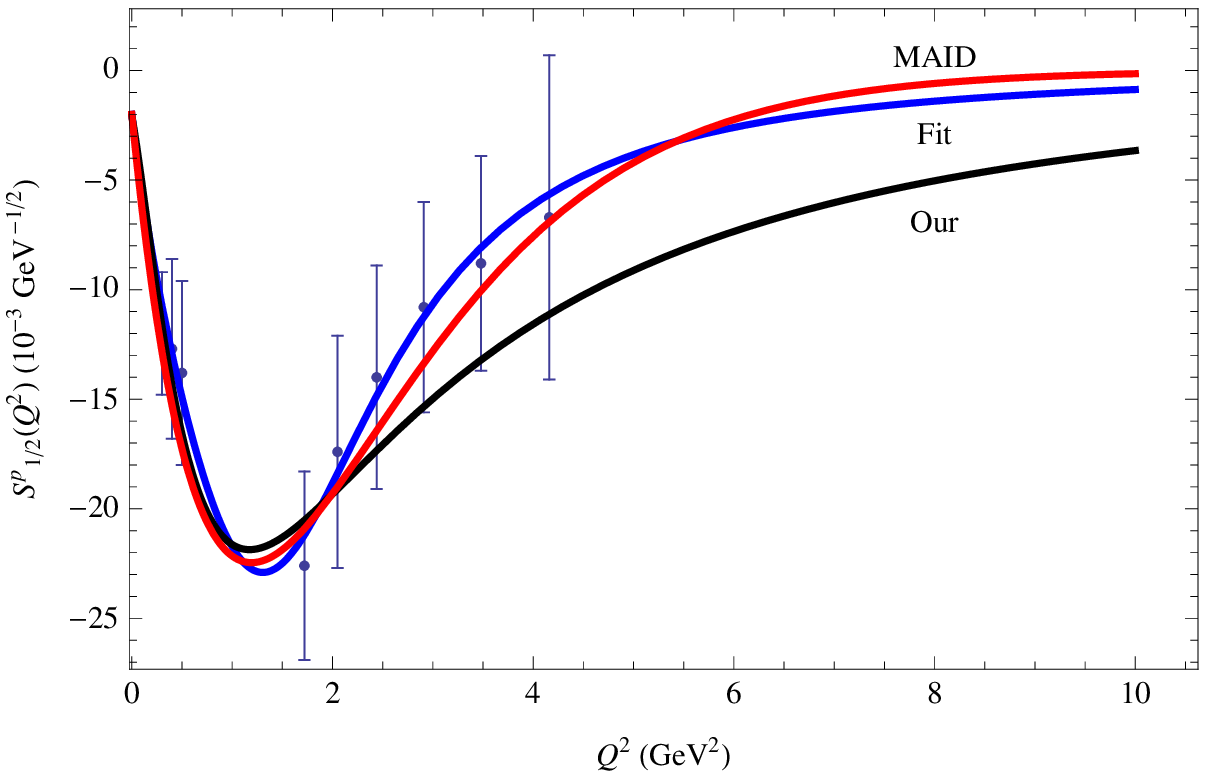,scale=.7}
\epsfig{figure=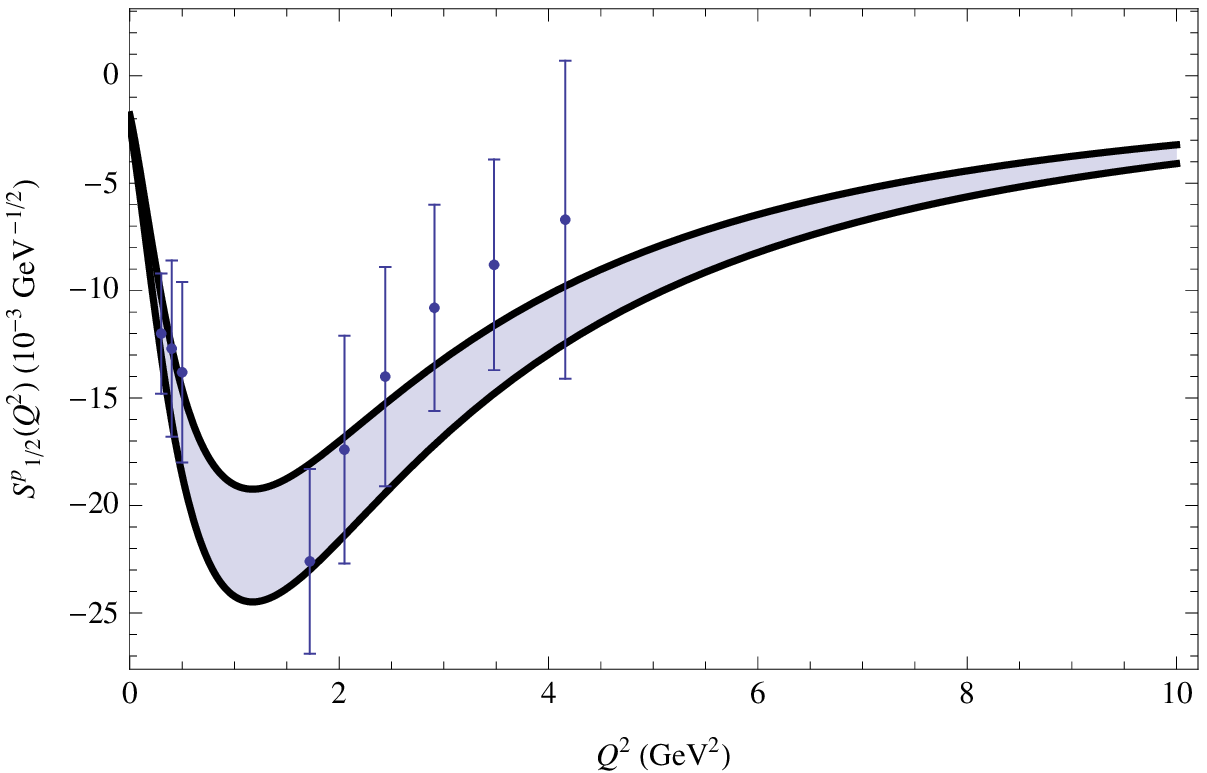,scale=.7}
\end{center}
\noindent
\caption{Helicity amplitude $S_{1/2}^p(Q^2)$ up to $Q^2=$ 10 GeV$^2$ 
with data taken from CLAS. 
Left panel: Displayed are our results (Our), 
the MAID parametrization (MAID), and (Fit) -  the parametrization 
of Eq.~(25). Right panel: Our results are shown for 
the case of variation of parameters of our approach (shaded band) 
in comparison with data. 
\label{fig2}}
\end{figure}

Our results for the $Q^2$ dependence of the helicity amplitudes in the
$N-N^*(1535)$ transition (proton channel) 
are fully displayed in the left panels in Figs.~\ref{fig1} and~\ref{fig2}. 
We compare them to experimental results of the CLAS Collaboration 
(JLab)~\cite{Aznauryan:2009mx,Aznauryan:2012ec,Mokeev:2015lda} and to 
the MAID parametrization~\cite{Drechsel:2007if} 
\eq
A_{1/2}^p(Q^2) &=& 0.066 \ {\rm GeV}^{-1/2} \
(1 + 1.61 \ {\rm GeV}^{-2} Q^2) \, 
\exp[-0.70 \ {\rm GeV}^{-2} Q^2]\,,
\nonumber\\
S_{1/2}^p(Q^2) &=&
-0.002 \ {\rm GeV}^{-1/2} \
(1 + 23.9 \ {\rm GeV}^{-2}) \ 
\exp[-0.81 \ {\rm GeV}^{-2} Q^2]\,.
\en
We further display a parametrization proposed by us: 
\eq
A_{1/2}^p(Q^2) &=& A_{1/2}^p(0) \,
\frac{1 + a_1 Q^2}{1 + a_2 Q^2 + a_3 Q^4 + a_4 Q^6} \,,
\nonumber\\
S_{1/2}^p(Q^2) &=& S_{1/2}^p(0) \,
 \frac{1 + s_1 Q^2}{1 + s_2 Q^2 + s_3 Q^4 + s_4 Q^6} \,,
\en
where
\eq
A_{1/2}^p(0) =  0.090  \ {\rm GeV}^{-1/2}\,, \quad
S_{1/2}^p(0) = -0.002  \ {\rm GeV}^{-1/2}\,, \quad
\en
and
\eq
& &
a_1 =   3.066  \ {\rm GeV}^{-2}\,, \quad
a_2 =   2.965  \ {\rm GeV}^{-2}\,, \quad
a_3 =   0.889  \ {\rm GeV}^{-4}\,, \quad
a_4 =   0.127  \ {\rm GeV}^{-6}\,,\nonumber\\
& &
s_1 =  20.460   \ {\rm GeV}^{-2}\,, \quad
s_2 =   1.325   \ {\rm GeV}^{-2}\,, \quad
s_3 =  -0.900   \ {\rm GeV}^{-4}\,, \quad
s_4 =   0.550   \ {\rm GeV}^{-6}\,.
\en 
We also present an analysis of the error of our approach due to 
a variation of the parameters (up to 15\%) for both helicity amplitudes 
in the right panels of Figs.~\ref{fig1} and~\ref{fig2}. 

\section{Summary}

We extended our formalism based on a soft-wall AdS/QCD approach to the 
description of the $\gamma N \to N^*(1535)$ transition. 
We showed that inclusion of the minimal electromagnetic coupling 
of the nucleon and the $N^*(1535)$ resonance, based on the coupling of two 
fermion AdS fields with the same twist-dimension, is manifestly 
gauge invariant. It then results in
a satisfactory description of data on the helicity amplitudes even 
at small $Q^2$. 
The failure to reproduce the low-$Q^2$ behavior of these helicity amplitudes
was a long-standing problem of most theory descriptions, the present soft-wall
AdS/QCD approach can offer a solution.
In the future we plan to apply our formalism to the calculation 
of electromagnetic transitions between the nucleon and further high-spin 
resonances. 

\begin{acknowledgments}

This work was funded by
the Carl Zeiss Foundation under Project ``Kepler Center f\"ur Astro- und
Teilchenphysik: Hochsensitive Nachweistechnik zur Erforschung des
unsichtbaren Universums (Gz: 0653-2.8/581/2)'', 
by ``Verbundprojekt 05A2017 - CRESST-XENON: Direkte Suche nach Dunkler 
Materie mit XENON1T/nT und CRESST-III. Teilprojekt 1''
(F\"orderkennzeichen 05A17VTA)'', by CONICYT (Chile) under
Grants No. 7912010025, No. 1180232 and PIA/Basal FB0821, and 
by FONDECYT (Chile) under Grant No. 1191103. 

\end{acknowledgments}


\begin{thebibliography}{999}

\bibitem{Devenish:1975jd} 
  R.~C.~E.~Devenish, T.~S.~Eisenschitz, and J.~G.~Korner,
  Phys.\ Rev.\ D {\bf 14}, 3063 (1976).

\bibitem{Aznauryan:2011qj} 
  I.~G.~Aznauryan and V.~D.~Burkert,
  Prog.\ Part.\ Nucl.\ Phys.\  {\bf 67}, 1 (2012).

\bibitem{Aznauryan:2009mx}   
  I.~G.~Aznauryan {\it et al.}  (CLAS Collaboration),
  Phys.\ Rev.\ C {\bf 80}, 055203 (2009).

\bibitem{Aznauryan:2012ec}  
  I.~G.~Aznauryan and V.~D.~Burkert,
  Phys.\ Rev.\ C {\bf 85}, 055202 (2012).

\bibitem{Mokeev:2015lda} 
  V.~I.~Mokeev {\it et al.}, 
  Phys.\ Rev.\ C {\bf 93}, 025206 (2016).

\bibitem{Tiator:2011pw}
  L.~Tiator, D.~Drechsel, S.~S.~Kamalov, and M.~Vanderhaeghen,
  Eur.\ Phys.\ J.\ Special Topics {\bf 198}, 141 (2011).

\bibitem{Stajner:2017fmh} 
  S.~Stajner, P.~Achenbach, T.~Beranek, J.~Bericic, J.~C.~Bernauer, 
  D.~Bosnar, R.~Bohm, L.~Correa, A.~Denig {\it et al.},
  Phys.\ Rev.\ Lett.\  {\bf 119}, 022001 (2017).

\bibitem{Brodsky:1973kr} 
  S.~J.~Brodsky and G.~R.~Farrar,
  Phys.\ Rev.\ Lett.\  {\bf 31}, 1153 (1973); 
  V.~A.~Matveev, R.~M.~Muradian and A.~N.~Tavkhelidze,
  Lett.\ Nuovo Cimento\  {\bf 7}, 719 (1973).  

\bibitem{deTeramond:2011qp}
  G.~F.~de Teramond and S.~J.~Brodsky,
  AIP Conf.\ Proc.\  {\bf 1432}, 168 (2012). 

\bibitem{Gutsche:2012wb} 
  T.~Gutsche, V.~E.~Lyubovitskij, I.~Schmidt, and A.~Vega,
  Phys.\ Rev.\ D {\bf 87}, 016017 (2013).

\bibitem{Ramalho:2017pyc} 
  G.~Ramalho and D.~Melnikov,
  Phys.\ Rev.\ D {\bf 97}, 034037 (2018). 

\bibitem{Ramalho:2017muv} 
  G.~Ramalho,
  Phys.\ Rev.\ D {\bf 96}, 054021 (2017). 

\bibitem{Gutsche:2017lyu} 
  T.~Gutsche, V.~E.~Lyubovitskij, and I.~Schmidt,
  Phys.\ Rev.\ D {\bf 97}, 054011 (2018). 

\bibitem{Gutsche:2019jzh} 
  T.~Gutsche, V.~E.~Lyubovitskij, and I.~Schmidt,
  Nucl.\ Phys.\ {\bf B952}, 114934 (2020).  

\bibitem{Gutsche:2019blp} 
  T.~Gutsche, V.~E.~Lyubovitskij, I.~Schmidt, and A.~Y.~Trifonov,
  Phys.\ Rev.\ D {\bf 99}, 054030 (2019); 
  Phys.\ Rev.\ D {\bf 99}, 114023 (2019).

\bibitem{Konen:1989jp}
  W.~Konen and H.~J.~Weber,
  Phys.\ Rev.\ D {\bf 41}, 2201 (1990);   
  R.~H.~Stanley and H.~J.~Weber,
  Phys.\ Rev.\ C {\bf 52}, 435 (1995).

\bibitem{Aznauryan:2004jd} 
  I.~G.~Aznauryan, V.~D.~Burkert, H.~Egiyan, K.~Joo, R.~Minehart, 
  and L.~C.~Smith,
  Phys.\ Rev.\ C {\bf 71}, 015201 (2005). 

\bibitem{Drechsel:2007if} 
  D.~Drechsel, S.~S.~Kamalov, and L.~Tiator,
  Eur.\ Phys.\ J.\ A {\bf 34}, 69 (2007).

\bibitem{Jido:2007sm} 
  D.~Jido, M.~D\"oring, and E.~Oset,
  Phys.\ Rev.\ C {\bf 77}, 065207 (2008). 

\bibitem{Braun:2009jy}
  V.~M.~Braun {\it et al.},
  Phys.\ Rev.\ Lett.\  {\bf 103}, 072001 (2009); 
  I.~V.~Anikin, V.~M.~Braun, and N.~Offen,
  Phys.\ Rev.\ D {\bf 92}, 014018 (2015).

\bibitem{Aliev:2013dxa}
  T.~M.~Aliev and M.~Savci,
  Phys.\ Rev.\ D {\bf 88}, 056021 (2013);
  T.~M.~Aliev, T.~Barakat, and K.~Simsek,
   Phys.\ Rev.\ D {\bf 100}, 054030 (2019). 

\bibitem{Ramalho:2011ae}
  G.~Ramalho and M.~T.~Pena,
  Phys.\ Rev.\ D {\bf 84}, 033007 (2011).  

\bibitem{An:2008xk} 
  C.~S.~An and B.~S.~Zou,
  Eur.\ Phys.\ J.\ A {\bf 39}, 195 (2009);
  Chin.\ Phys.\ C {\bf 34}, 245 (2010).

\bibitem{Ramalho:2011fa}
  G.~Ramalho and K.~Tsushima, 
  Phys.\ Rev.\ D {\bf 84}, 051301(R) (2011);
  G.~Ramalho,
  Phys.\ Lett.\ B {\bf 759}, 126 (2016); 
  Phys.\ Rev.\ D {\bf 95}, 054008 (2017).

\bibitem{Obukhovsky:2013fpa} 
  I.~T.~Obukhovsky, A.~Faessler, T.~Gutsche, and V.~E.~Lyubovitskij,
  Phys.\ Rev.\ D {\bf 89}, 014032 (2014). 

\bibitem{Obukhovsky:2019aa}
  I.~T.~Obukhovsky, A.~Faessler, D.~K.~Fedorov, T.~Gutsche, 
  and V.~E.~Lyubovitskij,
  Phys.\ Rev.\ D {\bf 100}, 094013 (2019). 

\bibitem{Gutsche:2011vb} 
  T.~Gutsche, V.~E.~Lyubovitskij, I.~Schmidt, and A.~Vega,
  Phys.\ Rev.\ D {\bf 85}, 076003 (2012); 
  A.~Vega, I.~Schmidt, T.~Gutsche, and V.~E.~Lyubovitskij,
  Phys.\ Rev.\ D {\bf 83}, 036001 (2011). 

\bibitem{Gutsche:2012bp} 
  T.~Gutsche, V.~E.~Lyubovitskij, I.~Schmidt, and A.~Vega,
  Phys.\ Rev.\ D {\bf 86}, 036007 (2012). 
 
\bibitem{Grigoryan:2007my}
  H.~R.~Grigoryan and A.~V.~Radyushkin,
  Phys.\ Rev.\ D {\bf 76}, 095007 (2007).
 
\bibitem{PDG18}
  M.~Tanabashi {\it et al.} (Particle Data Group),
  Phys.\ Rev.\ D {\bf 98}, 030001 (2018).

\bibitem{Kadeer:2005aq}
  A.~Kadeer, J.~G.~K\"orner, and U.~Moosbrugger,
  Eur.\ Phys.\ J.\ C\  {\bf 59}, 27 (2009).

\bibitem{Faessler:2009xn}
  A.~Faessler, T.~Gutsche, M.~A.~Ivanov, J.~G.~K\"orner,  
  and V.~E.~Lyubovitskij,
  Phys.\ Rev.\ D {\bf 80}, 034025 (2009); 
  T.~Branz, A.~Faessler, T.~Gutsche, M.~A.~Ivanov,
  J.~G.~K\"orner, V.~E.~Lyubovitskij, and B.~Oexl,
  Phys.\ Rev.\ D {\bf 81}, 114036 (2010); 
  T.~Gutsche, M.~A.~Ivanov, J.~G.~K\"orner, V.~E.~Lyubovitskij, 
  V.~V.~Lyubushkin, and P.~Santorelli,
  Phys.\ Rev.\ D {\bf 96}, 013003 (2017).

\end{thebibliography}
\end{document}